\def\etal{{\it et al.\/}\ }

\def\lsim{~\rlap{$<$}{\lower 1.0ex\hbox{$\sim$}}}
\def\gsim{~\rlap{$>$}{\lower 1.0ex\hbox{$\sim$}}}
\def\void#1{{}}
\def\bi{\begin{itemize}}
\def\ei{\end{itemize}}
\def\bc{\begin{center}}
\def\ec{\end{center}}

\documentclass[runningheads]{cl2emult}

\usepackage{makeidx}  
\usepackage{graphicx} 
\usepackage{subeqnar} 
\usepackage{multicol} 
\usepackage{cropmark} 
\usepackage{eso}      
\makeindex            

\begin{document}
\title*{EIS - An Imaging Survey for VLT Science}
%
%
%
%
\titlerunning{ESO Imaging Survey}
%
\author{Luiz da Costa} 
\authorrunning{Luiz da Costa}
%
%
\institute{European Southern Observatory, Garching bei M\"unchen,
D-85748, Germany}

\maketitle              

\begin{abstract} The imaging data assembled by the recently completed
ESO Imaging Survey (EIS) are reviewed and their scientific value
briefly assessed. Among the various applications, the imaging data has
been used to build a large sample of candidate distant clusters of
galaxies in the Southern Hemisphere to be used for follow-up
observations with the VLT as well as other space and ground-based
facilities. Preliminary results from ongoing work to confirm these
candidates are reported and the future prospects discussed.

\end{abstract}

\section{Introduction}

The ESO Imaging Survey (EIS) was a joint effort of ESO and its
community to carry out a public imaging survey and to prepare target
lists for different scientific applications in preparation for the
beginning of regular operation of the VLT. The primary scientific
goals were conceived by a Working Group (WG) composed of experts
selected from the community at large and the survey was conducted by a
team formed primarily by astronomers from the community.  Among the
different goals of EIS, one of the top priorities was the construction
of a catalog of distant cluster of galaxies candidates over an
extended redshift baseline from which targets could be drawn for
subsequent follow-up work. Such a sample was constructed by the EIS
team and it is the main focus of this presentation.

The importance of a large sample of confirmed distant clusters is
vast, as emphasized throughout this meeting, ranging from the study of
galaxy and large-scale structure formation and evolution to estimates
of cosmological parameters.  This has motivated several recent
attempts to find and investigate intermediate and high-redshift
clusters using $X$--ray, optically- and infrared-selected samples.
Even though $X$--ray samples are in many respects superior to those
selected in optical or infrared, the number of confirmed distant
clusters in the Southern Hemisphere is still quite small. Currently,
the best samples of distant $X$--ray clusters are all based on deep
pointings of the ROSAT PSPC~\cite{rosati2}.  In particular, the
RDCS~\cite{rosati1} includes about 30 $z> 0.5$ spectroscopically
confirmed clusters, the majority in the Northern Hemisphere.
Moreover, considering the currently foreseen timetable of future
$X$-ray missions it is clear that no new $X$-ray based high-redshift
cluster sample will be available in the next few years. Clearly, if
VLT is to play a role in the study of the distant clusters it will
have to rely on optically- or infrared--selected cluster samples, at
least in the near future. It is worth pointing out that while some
investigations require well-defined complete samples, usually
difficult to asses from optically-selected ones, others can benefit
solely by a statistically large sample.

The need for cluster targets was foreseen by the EIS WG which
recommended a moderately deep, wide-angle $I$-band survey to search
for distant clusters, despite the inadequacy of the available
detectors for wide-field imaging at the beginning of the
survey. Still, EIS is currently the largest imaging survey available
(17 square degrees) in the Southern Hemisphere to a depth ($I\lsim
23$) suitable for blind searches of distant clusters.

\section{General Results from EIS}

EIS was conceived as an attempt to optimize the use of the NTT by
carrying out a coordinated imaging survey in the one-year period
between the re-commissioning of the NTT and the commissioning of the
first unit of the VLT project. The main objectives of EIS were: 1) to
conduct a public optical-IR imaging survey in preparation for VLT; 2)
to produce target lists for VLT before the start of the Science
Verification period. The survey was designed to satisfy the
requirements of a broad range of applications. Following the
recommendations of the EIS WG the survey consisted of two parts: {\bf
1)} An optical, moderately deep, multi-passband, wide-angle survey
(EIS-WIDE) covering four patches distributed in the right ascension
range $22^h<\alpha<~10^h$ to produce targets almost year-round.  The
survey consisted of a mosaic of overlapping EMMI-NTT frames with each
position on the sky being sampled twice for a total integration time
of 300 sec.  The observations were carried out in the period July 97-
March 98, with all of the data, including pixel maps and derived
catalogs, being publicly released in March and July 1998; {\bf 2)}
Deep optical/infrared observations of the HDF-S and AXAF Deep Field
(EIS-DEEP). These observations were obtained using SUSI2 and SOFI at
the NTT in the period August-November 1998, with all of the data,
including derived catalogs, being publicly released December 10-12
1998~\cite{hdfs}~\cite{axaf}.

The accumulated imaging data provided the following data sets: {\bf
1)} $\sim$ 17 square degrees in $I$-band ($I_{AB}\sim 23.5$); {\bf 2})
$\sim$ 3 square degrees in $VI$; {\bf 3)} $\sim$ 2 square degrees in
$BVI$; {\bf 4)} $\sim$ 125 square arcmin in $JKs$ ($K_{AB}\sim 23.5$);
{\bf 5)} $\sim$ 80 square arcmin in $UBVRI$ ($I_{AB}\sim 26$). In
addition, combined, the optical/infrared deep observations provide a
coverage of 15 and 25 square arcmin in eight and seven passbands,
respectively~\cite{hdfs}~\cite{axaf}. Similar data will soon become
available for the HDF-S STIS field.

From the available photometric data several target lists were
extracted for immediate use in follow-up observations at different
facilities and for Science Verification observations. Among them are
candidate galaxy clusters, quasars, high-redshift galaxies as well as
galactic objects such as white dwarf and brown dwarf
candidates~\cite{zaggia}.  Some of these targets have already been
used for follow-up observations.  Quasar candidates identified in
patch B have been observed with 2dF at AAT, leading to the
confirmation of about 50\% of the candidates. Distant EIS cluster
candidates and high-redshift galaxies, identified using dropout and
photometric redshift techniques, have been successfully confirmed from
VLT observations using the VLT test camera~\cite{dacosta} and FORS
(see ESO PR 12/99), respectively, as part of the Science Verification
of UT1.

\section{EIS Cluster of Galaxies Candidates}

As part of the main effort of the EIS team, the galaxy catalogs
derived from the $I$-band images were used to search for candidate
clusters using the matched--filter algorithm~\cite{postman} to detect
clusters and to estimate redshifts.  The final EIS candidate clusters
sample consists of 304 objects in the redshift range $0.2\lsim
z\lsim1.3$~\cite{olsena}~\cite{olsenb}~\cite{marco}, with a median
redshift of $z\sim$0.5. This EIS cluster sample is the largest such a
catalog presently available in the Southern Hemisphere and it is
arguably one of the main results of the EIS-WIDE effort.

An extensive effort is now underway to validate these candidates at
different redshift intervals either by direct spectroscopy, using 4-m
class telescopes for the low-redshift ($z\lsim 0.5$) cluster
candidates, or by using optical/infrared color-magnitude relations to
identify the red-sequence of early-type cluster galaxies known to
exist over the redshift range considered. The latter has been carried
out by complementing the $I$-band EIS data with suitably deep optical
data ($z\lsim0.7$) and by infrared observations of high-redshift
candidates ($z>0.7$). The identification of the red-sequence is a
necessary step for a preliminary ``confirmation'' and for the
selection of targets for subsequent spectroscopic observations with
8-m class telescopes.

These follow-up observations include: {\bf 1)} 2dF@AAT test runs of
EIS cluster candidates in patch B and D~\cite{saglia}; {\bf 2)}
EFOSC2@3.6m observations of seven $z\sim 0.5$ clusters~\cite{ramella};
{\bf 3)} $JKs$ SOFI@NTT observations of fifteen clusters with
$z>0.6$~\cite{danish}~\cite{ramella}; {\bf 4)} deep optical
observations of two high-$z$ candidates with the
VLT-TC~\cite{dacosta}; {\bf 5)} deep optical observations ($VI$) of
four clusters with the Danish 1.5m telescope~\cite{danish}; and {\bf
6)} deep optical observations ($V$) of thirteen clusters with the
Nordic Optical Telescope~\cite{danish}.  In addition, deep $I-$band
imaging of $\sim 10$ clusters with $z\lsim0.5$ will be carried out in the
near future for weak lensing studies~\cite{seitz}. Altogether this
implies that about one-third of the EIS cluster candidates will have
already been followed-up by the end of 1999.

Gathering the information already available from the various groups,
the results from these preliminary confirmation efforts can be
summarized as follows: {\bf 1)} the combination of new spectroscopic
measurements, redshift data from the literature, and the presence of
color-magnitude relations~\cite{olsenb} provide either direct or
indirect evidence that most EIS clusters with $z<0.3$ are likely to be
real. Particularly promising are the preliminary results of the 2dF
observations for candidates with $z<0.5$. Test observations (3 hours
long), conducted under poor seeing conditions, showed that at least
50\% of the observed candidates have two or more concordant redshifts
and are very likely real associations; {\bf 2)} new imaging and
spectroscopic data for ten candidates with $0.4<z<0.6$ provide strong
evidence that nine of them are real associations; {\bf 3)} optical and
infrared color information for fifteen candidate clusters with $z>0.6$
suggest that twelve of them are real, including five at $z\gsim0.9$.

These preliminary results are extremely encouraging, suggesting that a
confirmation rate of about 70\% can be achieved. This clearly
justifies vigorously pursuing the confirmation effort. The data
already obtained have also shown ways to improve the confirmation
process.

\section {Future Prospects}

\subsection{Clusters}

The work on the EIS clusters is expected to proceed in several fronts
which include imaging and spectroscopic confirmation, velocity
dispersion measurements, deep imaging for weak lensing studies, and
$X-$ray observations with XMM. Such a program requires a variety of
observations that can be conducted at different facilities. For
instance, the detection of the early-type red sequence of intermediate
redshift candidates ($z\lsim0.7$) is possible from optical
observations with 2-m class telescopes, while for candidates at higher
redshifts infrared imaging is needed. Spectroscopic confirmation of
low-redshift EIS candidates ($z\lsim0.5$) can be successfully carried
out using 2dF@AAT. In fact, by the appropriate selection of galaxies
using color information, or whenever possible photometric redshifts,
as well as further constraints on the positioning of fibers will
certainly increase the success rate of confirmation, and can lead not
only to accurate redshift measurements but may also provide a robust
determination of the velocity dispersion.

Beyond $z>0.5$ 8-m class telescopes are required if one expects to
build a large sample of confirmed clusters in a timely fashion. To
confirm the reality of these candidates requires the observation of at
least a few galaxies and the best candidates are those lying along the
red-sequence. Even though FORS1 is not the ideal spectrograph, given
the geometry of the slits, the observations can be made more efficient
by using photometric redshift techniques. Using the optical/infrared
data accumulated in the process of determining the red-sequence and
those available from the public surveys one can select galaxies over
the field-of-view of FORS1 within a small estimated redshift interval
about the nominal cluster redshift. The effectiveness of the method
has been demonstrated for the two high-redshift candidates observed
with the VLT-TC for which $BVIJKs$ color catalogs are
available~\cite{arnouts}. By selecting galaxies in a photometric
redshift bin of 0.2 around the nominal cluster redshift the surface
density of galaxies in the field is reduced by a factor of more than
five, thereby significantly increasing the chances of targeting other
cluster members well-beyond the cluster core and with different
SEDs. It remains to be seen in practice how large the success rate of
measuring redshifts for cluster members will be using this
method. However, these observations should provide enough data to
confirm the reality of the cluster and, perhaps, to obtain a first
estimate of the velocity dispersion.  More accurate velocity
dispersion measurements will require the next generation of VLT
instruments (FORS2 and especially the integral field unit of VIMOS),
which would greatly benefit from having already a sample of confirmed
clusters. It is worth mentioning that the infrared data used to detect
red-sequences will also be used to search for higher redshift
clusters. Preliminary analysis of the $JKs$ data available, covering
about 500 square arcmin, have led to the discovery of at least seven
concentrations of red objects with estimated redshifts $1.1<z< 1.7$,
reminiscent of recently confirmed clusters at
$z\sim1.2$~\cite{rosati3}~\cite{stanford}. Finally, deep $I$-band
imaging is required to estimate weak lensing masses. For low-redshift
clusters this will require observations with the WFI@2.2, while at
higher redshift VLT observations are needed.

If the program outlined above is successful it will naturally provide
a unique data set of confirmed high-z clusters, with virial and weak
lensing masses, for follow-up observations with XMM in time for the
second announcement of opportunities.  This concerted study of
clusters over such an extended look-back time will provide a unique
database to investigate the evolution of clusters and their galaxy
population.

\subsection{Public Surveys}

Recently, the WG for public surveys endorsed the continuation of such
programs using the recently commissioned wide-field imager (WFI@2.2)
and the NTT for infrared observations. These surveys are expected to
be conducted by both the EIS team and outside groups under the same
general guidelines adopted for EIS. The two-year plan recommended by
the WG envisions a deep five-passband multicolor survey of three
square-degrees, complemented by deep infrared observations ($JKs$) of
an area of 1000 square arcmin. It also calls for shallower observation
of selected stellar fields in preparation for follow-up observations
with the FLAMES spectrograph. In the short-term the WG has also
reaffirmed its support for completing the EIS-wide survey in two
additional passbands ($BV$) of the 17 square degrees area covered in
$I-$band by EIS.

\section{Summary}

The initiative taken by ESO and the Member States to carry out a
coordinated public imaging survey in a one-year time-scale has been
successful providing a vast set of imaging data, for a wide-range of
applications, in time for the Science Verification period of UT1 and
for the first year of VLT operation. As a by-product the EIS
experiment has led to the development of a powerful data reduction
pipeline prototype and a survey environment, both essential
ingredients for the continuation of other Public Surveys using the
WFI@2.2 imager.  The target lists prepared by EIS have already been
used in a number of follow-up works, in preparation for VLT programs,
and for the Science Verification period, thus meeting the primary
requirement of the survey.  In particular, a significant effort has
been made by different groups to investigate the EIS candidate cluster
catalog leading to encouraging results which, if vigorously pursued,
could lead to the construction of a large sample of high-redshift
clusters for observations with the VLT as well as other facilities
such as HST and XMM.

The response of the community to EIS has been good, judging by the
interest manifested by the number of accesses to the EIS web pages,
the amount of data retrieved, the number of follow-up programs
undertaken over the past six months and the desire of the WG to
continue public surveys. However, a significant effort must still be
made to cope with the large volume of data expected from the
wide-field imager.

\medskip

\noindent {\bf Acknowledgments}

I would like to thank all of the past and present EIS team members for
their enormous effort and dedication in making it possible to meet the
deadlines of the survey. I would also like to thank M. Colless, H.
J{\o}rgensen, M. Ramella, R. Saglia and S. Cristiani for reporting
their preliminary results. Special thanks to M. Scodeggio, L. Olsen,
R.  Rengelink and M. Nonino for their contribution to the cluster work
reported here and A. Renzini for his support throughout this
program. Finally, I thank R. Giacconi for providing the coordinates
for the AXAF Deep Field.

\clearpage
\addcontentsline{toc}{section}{Index}
\flushbottom
\printindex

\end{document}